\documentclass[fleqn,usenatbib]{mnras}

\usepackage[T1]{fontenc}
\usepackage{ae,aecompl}

\usepackage{graphicx}	
\usepackage{amsmath}	
\usepackage{mathrsfs} 
\usepackage{amsfonts}
\usepackage{amssymb}	
\usepackage{threeparttable}
\usepackage{subfigure}
\usepackage{multirow}
\usepackage{multicol}
\usepackage{booktabs}
\usepackage{makecell}
\usepackage{float}

\title[Torque reversal of OAO 1657-415]{Torque reversal and orbital profile of X-ray pulsar OAO 1657-415}

\author[Z. Liao et al.]{
Zhenxuan Liao$^{1,2}$,
Jiren Liu$^{3}$\thanks{Email: liujiren@bjp.org.cn},
Peter A. Jenke$^{4}$,
Lijun Gou$^{1,2}$
\\
$^{1}$Key Laboratory for Computational Astrophysics, National Astronomical Observatory, Chinese Academy of Sciences, \\~~Datun Road 20A, Beijing 100012, People's Republic of China\\
$^{2}$School of Astronomy and Space Science, University of Chinese Academy of Sciences, Beijing 100049, People's Republic of China\\
$^{3}$Beijing Planetarium, Xizhimenwai Road, Beijing 100044, China\\
$^{4}$University of Alabama in Huntsville, Huntsville, AL 35812, USA
}

\date{Accepted XXX. Received YYY; in original form ZZZ}

\pubyear{2021}

\begin{document}
\label{firstpage}
\pagerange{\pageref{firstpage}--\pageref{lastpage}}
\maketitle

\begin{abstract}
OAO 1657-415 is an atypical supergiant X-ray binary among wind-fed and disk-fed systems, showing alternate spin-up/spin-down intervals lasting on the order of tens of days. 
We study different torque states of OAO 1657-415 based on the spin history monitored by {\it Fermi}/GBM, together with fluxes from {\it Swift}/BAT and {\it MAXI}/GSC. 
Its spin frequency derivatives are well correlated with {\it Swift}/BAT fluxes during rapid spin-up episodes, anti-correlated with {\it Swift}/BAT fluxes during rapid spin-down episodes, and not correlated in between.
The orbital profile of spin-down episodes is reduced by a factor of 2 around orbital phases of 0.2 and 0.8 compared to that of spin-up episodes.
The orbital hardness ratio profile of spin-down episodes is also lower than that of spin-up episodes around phases close to the mid-eclipse, implying that there is more material between the neutron star and the observer for spin-down episodes than for spin-up episodes around these phases.
These results indicate that the torque state of the neutron star is connected with the material flow on orbital scale and support the retrograde/prograde disk accretion scenario for spin-down/spin-up torque reversal.

\end{abstract}




\section{Introduction}
\label{sec:intro}
In high mass X-ray binaries (HMXBs), a compact object, often a neutron star, accretes material from a massive optical star. 
According to the type of optical companion, HMXBs can be classified into supergiant X-ray binaries (SgXBs) and Be X-ray binaries (BeXBs).
In BeXBs the neutron star accretes from an equatorial decretion disk formed due to the rapid rotation of Oe/Be star,
while in SgXBs the accreted material is usually from the massive stellar wind of the optical companion \citep[for a recent review, see][]{K2020HMXBrv}. 
Wind accretion is generally evaluated with quasi-spherical Bondi-Hoyle-Lyttleton model, in which the transfer of angular momentum is negligible. 
Transient disks may be formed due to the spatial gradient of wind density and velocity \citep{SL76,Wang81}. 
Recently, it was found that when the wind velocity is comparable to the orbital velocity of the neutron star, the wind could be significantly beamed leading to the formation of a disk-like structure \citep{Mellah2019wcdisk,Karino19,Xu2019}. 
In a few SgXB systems (Cen X-3, SMC X-1, and LMC X-4), the optical companion is filling its Roche lobe, and the compact star is disk-fed through Roche-lobe overflow.

Unlike quasi-spherical wind accretion, disk accretion can efficiently transfer angular momentum to the neutron star and change its spin period, which can be easily measured on a timescale of days due to the small moment of inertia of neutron stars. 
Occasional spin-up episodes related to transient accretion disks have been reported in some wind-fed systems, such as GX 301-2 \citep{Koh1997,Liu2020}.
On the other hand, many X-ray pulsars were found to show spin-up/spin-down reversals on timescale from days to decades, such as Cen X-3, Vela X-1, OAO 1657-415, GX 1+4, and 4U 1626-67, which is a puzzling phenomenon not well understood yet \citep[e.g.][]{Bild1997,Chak1997,Camero2010}.

\begin{figure}
    \flushright
    \includegraphics[width=1.1\columnwidth]{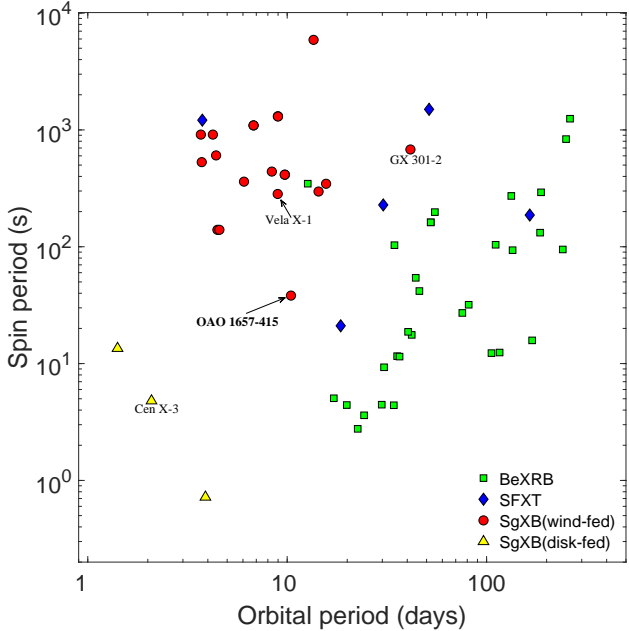}
    \caption{Orbital periods and spin periods of HMXBs plotted on the Corbet diagram. 
    Red circles represent wind-fed SgXBs, yellow triangles for disk-fed SgXBs, 
    green squares for BeXBs, and blue diamonds for Supergiant Fast X-ray Transients (SFXTs). 
    In this diagram, OAO 1657-415 occupies a particular position away from both disk-fed and wind-fed SgXBs, indicating that OAO 1657-415 is in an intermediate state. 
    }
    \label{fig:pcd}
\end{figure}

\begin{figure*}
    \flushleft
    \includegraphics[width=1.05\textwidth]{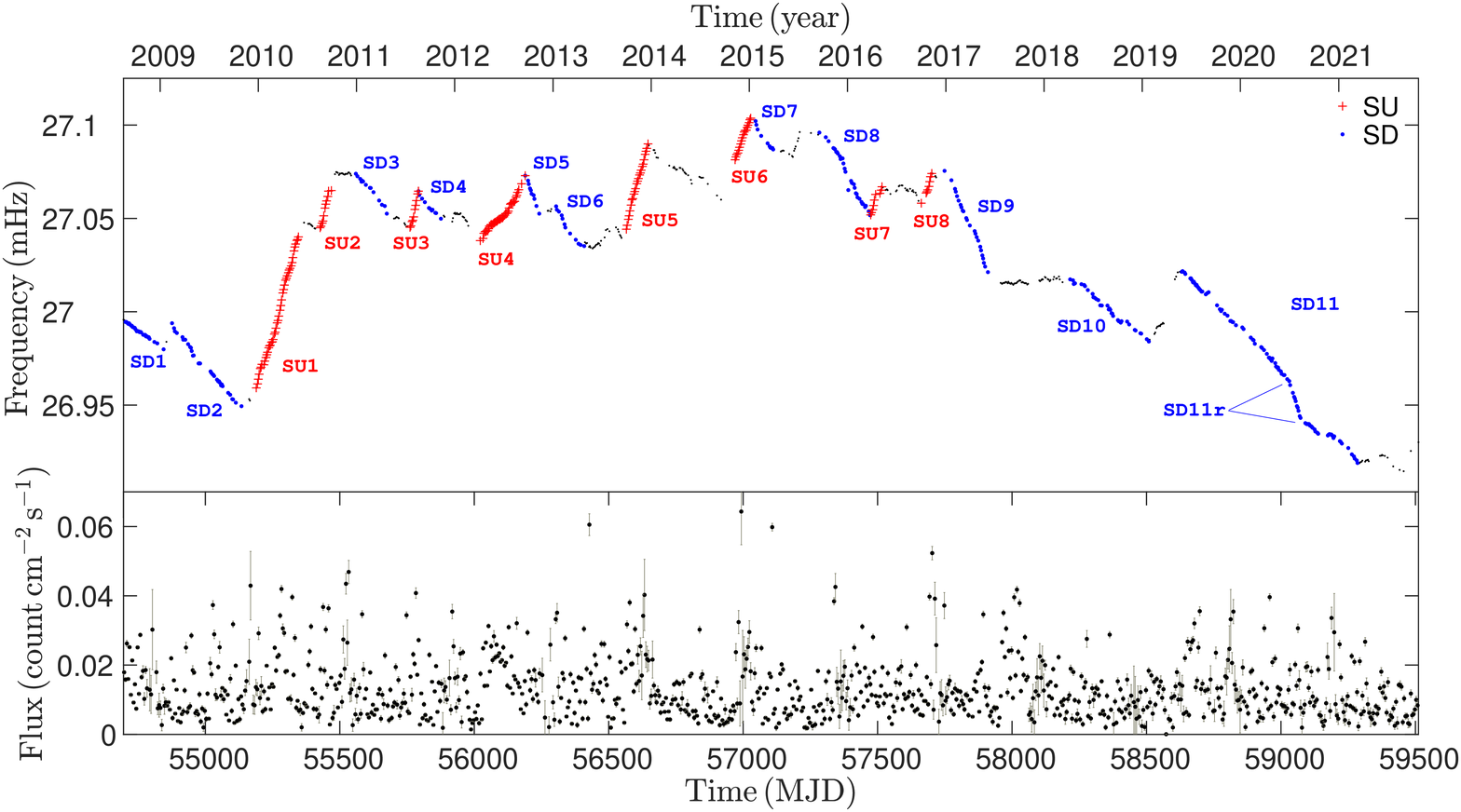}
    \caption{Upper panel: decade-long spin frequency history of OAO 1657-415 monitored by {\it Fermi}/GBM during MJD 54695--59515; The marked segments with numbers are intervals to be analyzed. 
The errors of spin frequency measurements are on the order of $10^{-7}\,\rm Hz$, too small to be displayed.
	 Lower panel: the flux within 15-50\,keV band extracted from {\it Swift}/BAT, binned in 5\,days.}
    \label{fig:mki}
\end{figure*}

The different types of HMXBs occupy different positions in the Corbet diagram, 
where the spin period is plotted against the orbital period \citep{Corbet1986}. 
The eclipsing X-ray binary OAO 1657-415 occupies a special position in the Corbet diagram, away from both wind-fed and disk-fed SgXBs, suggesting that it may represent a transition state bridging them (Fig.~\ref{fig:pcd}). 
It has a pulsation period of 38.22\,s initially measured by {\it HEAO}-1 \citep{White1979}, an orbital period of $\sim10.4$ days, and an eccentricity of 0.1075 \citep[e.g.][]{Chak1993,Jenke2012}. 
The optical companion of OAO 1657-415 (with a mass $\sim14.3\pm0.8\,M_{\sun}$) is classified as an Ofpe/WNL star with a hydrogen-depleted atmosphere, which is in a transitional stage towards a Wolf-Rayet star and much more evolved than typical optical companions in SgXBs \citep{Mason2009, Mason2012}. 

The early spin history of OAO 1657-415 showed a long-term trend of spin-up ($\nu/\dot{\nu}\sim1000\,\rm yr$) superimposed with alternate spin-up and spin-down periods as monitored by BATSE onboard the Compton Gamma Ray Observatory ({\it CGRO}) \citep{Bild1997,Barn2008}. 
The spin history of OAO 1657-415 monitored by {\it Fermi}/GBM over the last 13 years showed a spin-up trend before 2011, wandering variations between 2011 and 2017, and then a rare and relatively long spin-down trend after 2017 (Fig.~\ref{fig:mki}). 
\citet{Jenke2012} found that, during the GBM-monitored spin-up episodes, its spin frequency derivatives ($\dot{\nu}$) are well correlated with {\it Swift}/BAT fluxes, suggesting the existence of stable prograde accretion disk, while during the spin-down episodes, the non-correlation between $\dot{\nu}$ and fluxes may indicate wind accretion.

\begin{figure}
    \flushleft
    \includegraphics[width=1.1\columnwidth]{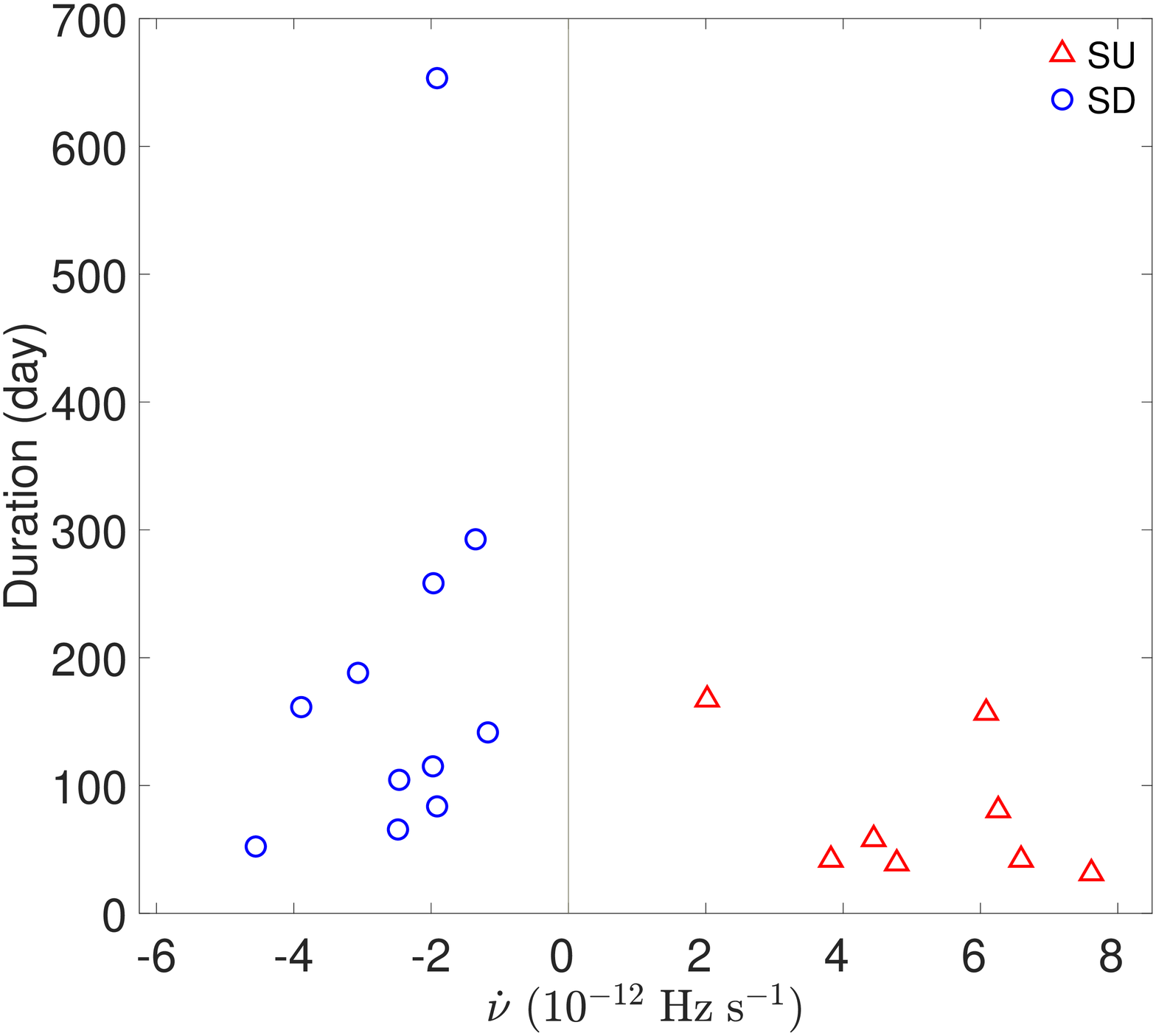}
    \caption{$\dot{\nu}$ versus duration for each interval.
    In general, spin-up intervals show shorter durations and higher spin changing rates, compared with spin-down intervals. }
    \label{fig:dnVdu}
\end{figure}

To explain the short-term torque reversals of OAO 1657-415, \citet{Bay1997} suggested episodic formation of prograde and retrograde accretion disks in the context of stellar wind accretion, but no correlation between $\dot{\nu}$ and fluxes was found based on {\it CGRO}/BATSE data. 
In this paper, we study the temporal and spectral properties of OAO 1657-415 during spin-up and spin-down episodes monitored by {\it Fermi}/GBM, with the aim to further understand the phenomenon of torque reversals between spin-up and spin-down.

\section{Observations}
\label{sec:obs}

The Gamma Ray Burst monitor \citep[GBM,][]{Meegan2009} onboard the {\it Fermi} Gamma-ray Space Telescope was launched in 2008. 
It has 4 clusters of 3 NaI detectors sensitive to 8-1000 keV on each corner oriented in different directions.
There are three public data types available with GBM: CTIME data which has 0.256\,s resolution and is available in 8 energy channels, 
CSPEC data which has 4.096\,s resolution with 128 energy channels and is used for spectroscopy, 
and CTTE data which is time tagged event data containing arrival times (2$\,\mu$s accuracy) and energy channels (0-127) of photons.
The GBM Accreting Pulsar Program (GAPP) monitors 44 accreting pulsars using CTIME data and publishes frequency and pulsed flux histories on its webpage\footnote{http://gammaray.msfc.nasa.gov/gbm/science/pulsars}.
For OAO 1657-415, the eclipsing time was excluded and the non-eclipsing time within one orbital period ($\sim9$\,days) was divided into 3 segments. 
The spin frequency is searched based on $Y_{\rm n}$ statistic, and the frequency with $Y_{\rm n}$ above a certain value is considered a detection \citep[for more details, see][]{Fin09,GBM10yrs}.

The Burst Alert Telescope (BAT) onboard the Neil Gehrels {\it Swift} Observatory has a large field of view and high sensitivity. It has been continuously monitoring the X-ray sky ranging from 15\,keV to 200\,keV on a daily basis since 2005 \citep{Krimm2013}. 
Daily and orbital light curves of hard X-ray transient program\footnote{https://swift.gsfc.nasa.gov/results/transients/BAT\_current.html\#anchor-EXO1657-419} are used in our data analysis. 

The Monitor of All-sky X-ray Image ({\it MAXI}) is the first high energy astrophysical experiment placed on the International Space Station since 2009. 
Its Gas Slit Camera \citep[GSC,][]{GSC} has been continuously monitoring the whole sky within 2-20\,keV band and providing daily and orbital light curves of transient sources\footnote{http://maxi.riken.jp/star\_data/J1700-416/J1700-416.html}.


\section{Data Analysis and results}
\label{sec:res}

\begin{table*}
    \renewcommand\arraystretch{1.5}
    \centering
	\caption{Duration and $\Dot{\nu}$ of each selected interval.}
    \begin{threeparttable}
    \begin{tabular}{cccccccc}
    \hline
    \#ID & Time  & Duration & $\Dot{\nu}$ & \#ID & Time & Duration & $\Dot{\nu}$ \\
    & (MJD) & (day) & $\rm (10^{-12} Hz\;s^{-1})$ & & (MJD) & (day) & $\rm (10^{-12} Hz\;s^{-1})$ \\
    \hline
    SU\,1 & 55189-55347 & 158 & 6.08(10) & SD\,1 & 54703-54845 & 142 & -1.17(02) \\
    SU\,2 & 55426-55469 & 43 & 6.59(50) & SD\,2 & 54876-55135 & 259 & -1.97(03) \\
    SU\,3 & 55761-55793 & 32 & 7.62(39) & SD\,3 & 55559-55675 & 116 & -1.97(08) \\
    SU\,4 & 56022-56190 & 168 & 2.02(07) & SD\,4 & 55792-55877 & 85 & -1.91(11) \\
    SU\,5 & 56565-56647 & 82 & 6.26(11) & SD\,5 & 56189-56242 & 53 & -4.55(18) \\
    SU\,6 & 56970-57029 & 59 & 4.45(10) & SD\,6 & 56304-56409 & 105 & -2.46(15) \\
    SU\,7 & 57474-57516 & 42 & 3.82(48) & SD\,7 & 57046-57112 & 66 & -2.48(16) \\
    SU\,8 & 57662-57702 & 40 & 4.78(49) & SD\,8 & 57283-57472 & 189 & -3.06(09) \\
    & & & & SD\,9 & 57748-57911 & 163 & -3.89(10) \\
    & & & & SD\,10 & 58216-58509 & 293 & -1.35(03) \\
    & & & & SD\,11 & 58631-59285 & 654 & -1.91(02) \\
    
    \hline
    \end{tabular}
    \end{threeparttable}
    \label{tab:nudot}
\end{table*}

OAO 1657-415 shows relatively smooth and continuous spin-up and spin-down episodes compared to other SgXBs, such as Vela X-1 \citep{L2020} and GX 301-2 \citep{Liu2020}. 
In order to study the differences between spin-up and spin-down episodes of OAO 1657-415, we divide the GBM-monitored long-term spin history into several spin-up (SU) and spin-down (SD) intervals, as illustrated in Fig.~\ref{fig:mki}. 
They are based on a continuous trend of spin-up or spin-down. The spin frequency change of each interval is required
to be larger than $1.4\times10^{-5}\,\rm Hz$, and the period of anomalous behavior in each interval (such as spin-up periods in spin-down interval, vice versa) should be less than 11\,days (about one orbital cycle).

For each interval, we calculate a mean spin changing rate ($\dot{\nu}$) by a linear fit.
The results are listed in Table.~\ref{tab:nudot}, and plotted in Fig.~\ref{fig:dnVdu}.
As can be seen, the absolute values of $\dot{\nu}$ of spin-up intervals are generally larger than those of spin-down intervals, except for SU\,4, which shows a small spin-up rate. 
Meanwhile, the durations of most spin-up intervals are shorter than 80\,days except for SU\,1 and SU\,4. 
In contrast, the durations of spin-down intervals are generally longer than, or close to, 80\,days. 
Combining the different spin changing rates and durations of spin-up and spin-down intervals, the spin frequencies of OAO 1657-415 over the decade-long GBM-monitored period show almost no net changes (Fig.~\ref{fig:mki}).

\begin{figure*}
    \centering
    \includegraphics[width=0.9\textwidth]{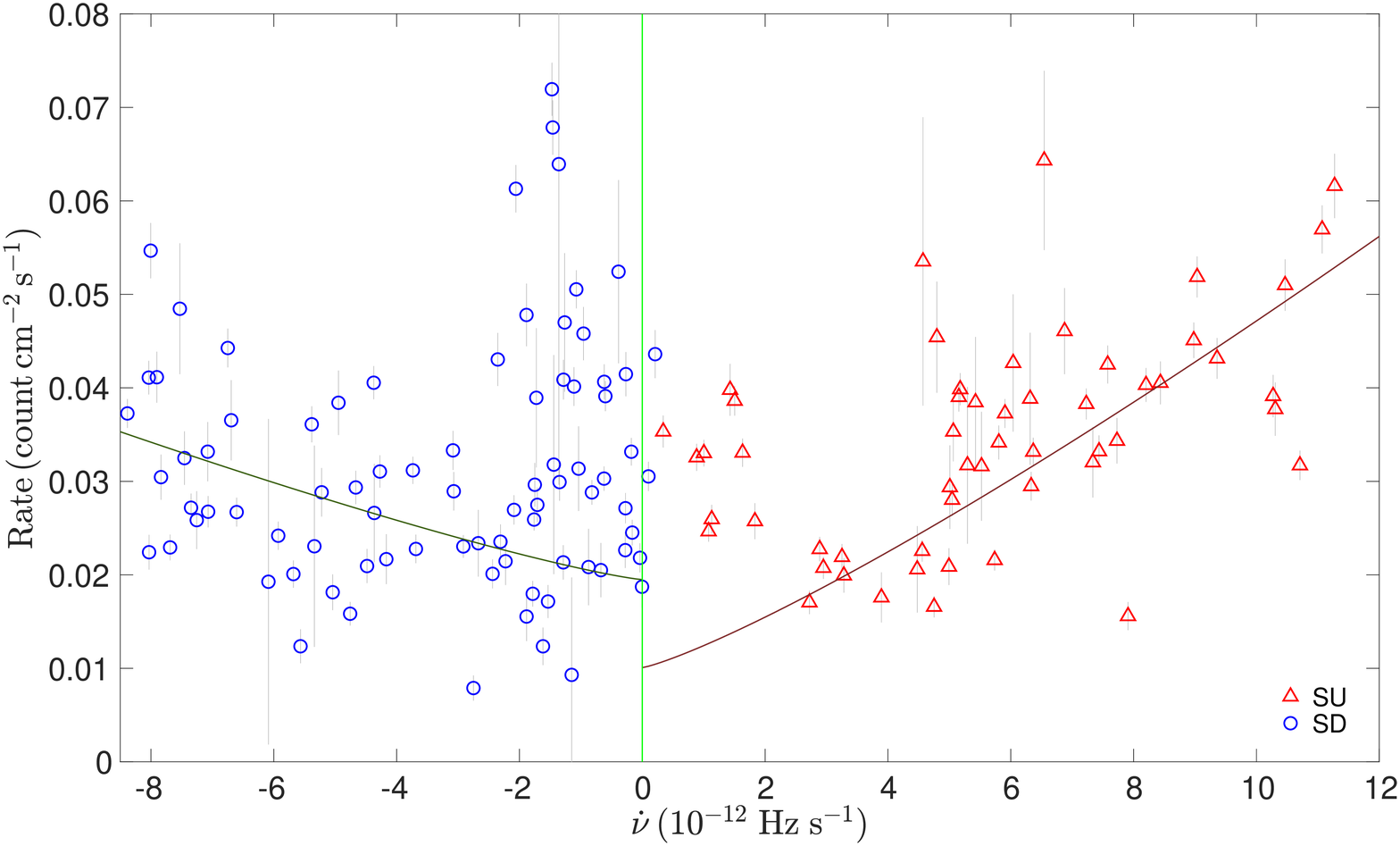}
    \caption{Correlation between spin frequency derivatives and {\it Swift}/BAT fluxes within 15-50\,keV band. 
    Red triangles are for spin-up intervals, while blue circles for spin-down intervals.
    The fitted curves for rapid ($|\dot{\nu}| \gtrsim 3\times10^{-12}\,\mathrm{Hz}\,\mathrm{s}^{-1}$) spin-up and
	 spin-down episodes are also plotted.
    }
    \label{fig:lvnd}
\end{figure*}

\subsection{Correlation between $\dot{\nu}$ and fluxes}
\label{sec:cor}

In the standard scenario of disk accretion, a correlation between accretion luminosity ($\mathscr{L}$) and torque exerted on the neutron star is expected to have the form: $\dot{\nu}\propto\mathscr{L}^{\frac{6}{7}}$ \citep{Rapp1977,GL1979b}.
Hence we compare $\dot{\nu}$ with {\it Swift}/BAT fluxes ($\mathscr{F}$) within selected spin-up/spin-down intervals.
To calculate $\dot{\nu}$, we only choose those orbital periods that have two or three spin detections, and $\dot{\nu}$ is calculated from two-point or three-point numerical derivative formula.
The fluxes are taken from the 15--50\,keV daily light curve monitored by {\it Swift}/BAT. 
Because OAO 1657-415 undergoes severe absorption \citep{Naik2009}, we take the maximum flux within one binary orbital period to represent the intrinsic flux, as \citet{Jenke2012} did. 
The calculated $\dot{\nu}$ and {\it Swift}/BAT fluxes within spin-up and spin-down intervals are plotted in Fig.~\ref{fig:lvnd}.

As can be seen, the distribution of $\dot{\nu}(\mathscr{F})$ can be roughly separated into three regions divided roughly by $\dot{\nu}$ boundaries at $-3$ and $ 3 \times 10^{-12} \,\rm Hz\,s^{-1} $: $\dot{\nu} \gtrsim 3\times10^{-12}$Hz\,s$^{-1}$, $\dot{\nu} \lesssim -3\times10^{-12}$Hz\,s$^{-1}$, and those in between. 
For spin-up intervals, only the interval of SU\,4 has spin rates of $\dot{\nu}<2\times10^{-12}$Hz\,s$^{-1}$, and the fluxes of all other spin-up intervals show a good correlation with $\dot{\nu}$.
For spin-down intervals, there are more intervals with $|\dot{\nu}|\lesssim 3\times10^{-12}$Hz\,s$^{-1}$, and their fluxes are scattered vertically, as those of SU\,4. 
On the other hand, the spin-down intervals of $\dot{\nu} \lesssim -3\times10^{-12}$Hz\,s$^{-1}$ show a negative correlation between $\dot{\nu}$ and fluxes, and the scattering seems a little larger for spin-down intervals than spin-up ones. 

We use a power-law function plus a constant to fit the relationship between $\dot{\nu}$ and fluxes separately for spin-up and spin-down intervals. 
For rapid spin-up intervals (excluding data points with $\dot{\nu} < 2\times10^{-12}$Hz\,s$^{-1}$, all of which come from SU\,4), we get $\dot{\nu}\propto\mathscr{F}^{0.84\pm0.26}$. 
On the other hand, for those data points in the rapid spin-down regime with $\dot{\nu} < -3\times10^{-12}$Hz\,s$^{-1}$, we get $\dot{\nu}\propto-\mathscr{F}^{0.83\pm0.43}$.

\subsection{Pulse profile}
\label{sec:pp}

\begin{figure}
    \centering
    \includegraphics[width=1.09\columnwidth]{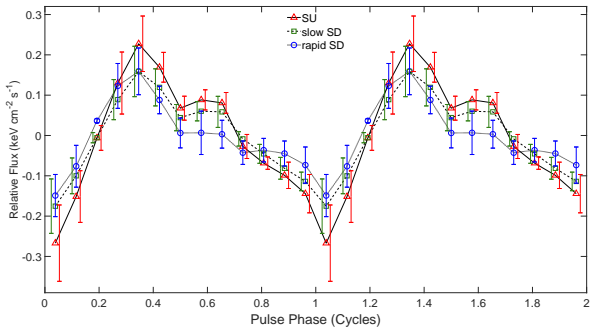}
    \caption{The pulsed profile within 12--25\,keV band for spin-up, slow spin-down, and rapid spin-down episodes, obtained from {\it Fermi}/GBM. 
    The sub-peak at phase 0.6 is less prominent for rapid spin-down than for spin-up and slow spin-down episodes. 
    For clarity, errorbars of spin-up episodes are shifted rightward, while errorbars of slow spin-down episodes are shifted leftward.
    }
    \label{fig:ppc1E}
\end{figure}

The pulse profile of OAO 1657-415 is energy-dependent, and shows a complex shape dominated by one main peak with a sub-peak that is $\sim\,$0.25 spin phase away from the main peak \citep[e.g.][]{Bild1997,Barn2008,Jenke2012,oao2021}. 
No attempt has been made to compare pulse profiles in different torque states.
We extract the pulse profile of OAO 1657-415 from spin-up, rapid spin-down, and slow spin-down episodes, separately.
There is a rapid spin-down period (MJD 59030-59073, $\dot{\nu}=-5.37(10)\times10^{-12}\rm\,Hz\,s^{-1}$) in SD 11, and we identify it as `SD\,11r'.
Based on the fitted $\dot{\nu}$ of each interval, SD\,5,9,11r are chosen as the representation of rapid spin-down episodes, while SD\,1,2,3,4,10 and the rest of SD\,11 for slow spin-down episodes. All the spin-up intervals are used for the extraction of pulse profile.
Since the {\it Fermi}/GBM data is dominated by the background, the constant component of the light curve has been subtracted and only the pulsed flux is obtained \citep[e.g.][]{Fin09}.
The averaged 12--25\,keV pulse profiles of different episodes are plotted in Fig.~\ref{fig:ppc1E}. 
The phases of different pulse profiles have been aligned with a cross-correlation method, and some exceptions are removed. The shapes of all profiles are similar in general. 
The amplitudes of pulse profiles of the spin-up episodes are larger than those of other episodes. 
We note that the sub-peak around the spin phase 0.6 of spin-up episodes is significant on 90\% confidence level, while the sub-peak of rapid spin-down episodes is much less prominent.

\subsection{Orbital profile}
\begin{figure*}
    \centering
    \subfigure[]{
    \includegraphics[width=1.7\columnwidth]{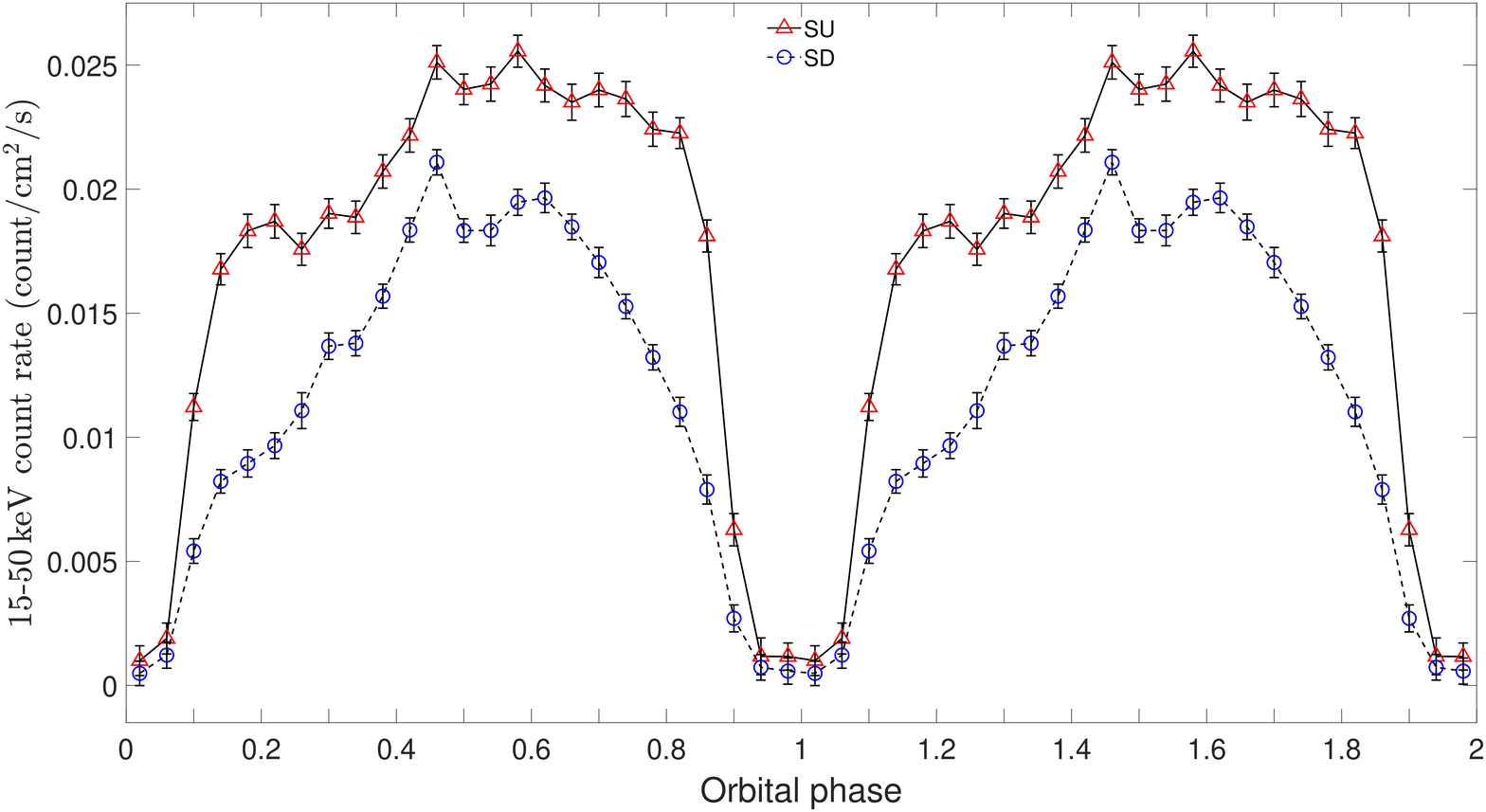}
    \label{fig:eopf}
    }
    \subfigure[]{
    \includegraphics[width=1.7\columnwidth]{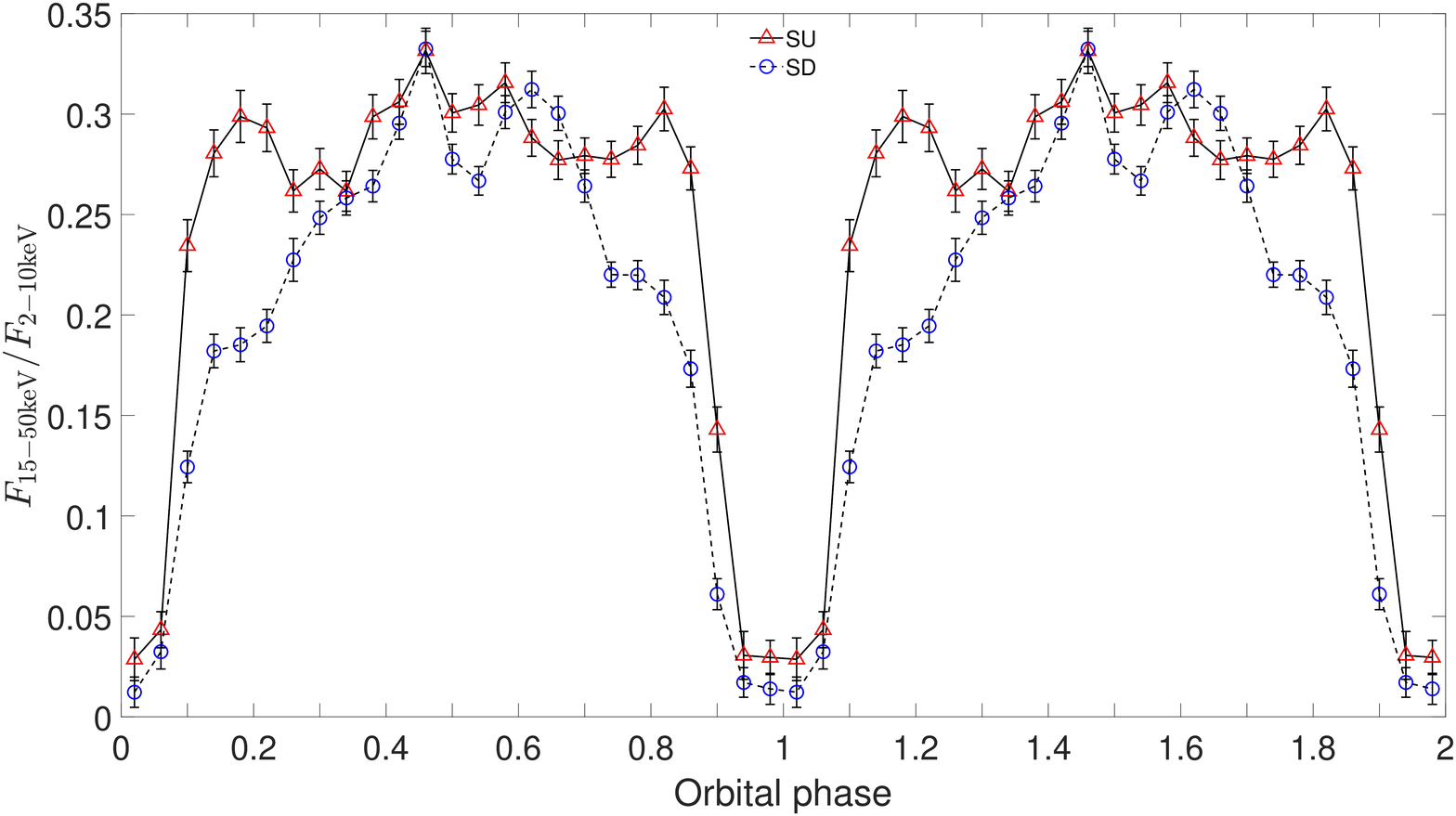}
    \label{fig:udhr}
    }
    \caption{Top: orbital profile of OAO 1657-415 extracted from {\it Swift}/BAT light curves within 15-50\,keV band for spin-up and spin-down episodes. Bottom: orbital hardness ratio profile, defined as the ratio of 15-50\,keV flux (from {\it Swift}/BAT) to 2-10\,keV flux (from {\it MAXI}/GSC), for spin-up and spin-down episodes. The HRs of spin-down episodes during ingress and egress periods are smaller than those of spin-up episodes.}
\end{figure*}

The orbital profile of OAO 1657-415 shows an asymmetric shape, with a broad peak around orbital phases of 0.5--0.8 \citep[e.g.][]{wen2006,Barn2008,oao2014suzaku,Fala2015}.
To study the possible orbital differences between spin-up and spin-down episodes, we extract the orbital profile from the {\it Swift}/BAT orbital light curves within 15-50\,keV band for spin-up, rapid spin-down, and slow spin-down episodes, separately. 
The orbital parameters of an orbital period of 10.44729 days, a period derivative of $-9.74\times10^{-8}\rm\,days/day$, and a phase zero time of MJD 52298.01 are adopted, where the zero orbital phase is defined at the mid-eclipsing point \citep{Jenke2012}.
Since we found no significant differences of the profiles between rapid and slow spin-down episodes, the profiles from rapid and slow spin-down episodes are combined together.
The results are presented in Fig.~\ref{fig:eopf}. 

The averaged fluxes of spin-down episodes are generally smaller than those of spin-up episodes. The orbital profile of spin-up episodes is peaked around phases of 0.5-0.8, similar to previous studies. 
The orbital profile of spin-down episodes, however, is more symmetric compared with that of spin-up episodes. 
Specifically, the fluxes of spin-down episodes around orbital phases of 0.2 and 0.8, which are just before the ingress and after the egress, are reduced by a factor of 2 compared with those of spin-up episodes.
The different orbital profiles of spin-up and spin-down episodes indicate that the material flow to the neutron star is different for spin-up and spin-down episodes.
Either the mass accretion rate ($\dot{M}$) of spin-down episodes is smaller and/or 
the emission of spin-down episodes is more obscured.

To further explore the orbital material distribution of spin-up and spin-down episodes, we fold the hardness ratio (HR), defined as $\mathscr{F}_{\rm 15-50\,keV}/\mathscr{F}_{\rm 2-10\,keV}$, where the former is obtained from {\it Swift}/BAT light curves and the latter from {\it MAXI}/GSC, for spin-up and spin-down episodes separately. 
As shown in Fig.~\ref{fig:udhr}, during the ingress phase, the HR becomes smaller when the orbital phase is closer to the mid-eclipse point. 
Since there is more material between the neutron star and the observer as the neutron star proceeds further during the ingress, it means that with more material, the {\it Swift}/BAT fluxes are obscured more heavily while the {\it MAXI}/GSC fluxes are less affected. 
This is the case when soft energy photons are mainly from scattered emission, instead of intrinsic emission.

On the other hand, at a certain ingress phase ($\sim\,$0.85--0.95), the HR of spin-down episodes is smaller than that of spin-up episodes. 
If the intrinsic spectrum is similar for both spin-up and spin-down episodes, it means that at a certain ingress phase, hard photons within the {\it Swift}/BAT band of spin-down episodes are more obscured than those of spin-up episodes. 
That is, at a certain ingress phase, there is more material between the neutron star and the observer for spin-down episodes than for spin-up episodes. 
The same behavior of HR is also happened during the egress phase.
This evidence supports the scenario that the relatively reduced fluxes of spin-down episodes around orbital phases of 0.2 and 0.8 (Fig.~\ref{fig:eopf}) are primarily due to stronger obscuration of spin-down episodes than spin-up episodes.     

\section{Discussion and conclusion}
\label{sec:dc}

We studied the temporal and spectral behavior of OAO 1657-415 during different torque states based on the spin history monitored by {\it Fermi}/GBM. 
The spin-up episodes generally have shorter duration and higher spin changing rates than spin-down episodes. 
Its spin frequency derivatives are well correlated with {\it Swift}/BAT fluxes during rapid spin-up episodes ($\dot{\nu} \gtrsim 3\times 10^{-12} \rm Hz \, s^{-1}$), anti-correlated with {\it Swift}/BAT fluxes during rapid spin-down episodes ($\dot{\nu} \lesssim -3\times 10^{-12} \rm Hz \, s^{-1}$), and not correlated in between.
The orbital profile of spin-down episodes is relatively reduced around orbital phases of 0.2 and 0.8 compared to that of spin-up episodes. 
The orbital hardness ratio profile of spin-down episodes is also smaller than that of spin-up episodes during ingress and egress periods, indicating that there is more material between the neutron star and the observer during ingress and egress periods for spin-down episodes than for spin-up episodes.


The anti-correlation between torque and luminosity was first reported by \citet{Chak1997} for GX 1+4 during a secular spin-down episode monitored by {\it CGRO}/BATSE. 
Such an anti-correlation is in conflict with the standard magnetic disk accretion theory, which predicts a monotonic relation between torque and luminosity that a higher mass accretion rate (hence higher luminosity) should yield a larger torque \citep{GL1979b}. 
The spin-down torque could be explained by a retrograde accretion disk, where the material is rotating in the opposite sense as the rotation of neutron star \citep{Maki88,Nelson1997,Chak1997}.

The anti-correlation between torque and flux during the rapid spin-down episodes of OAO 1657-415 provides additional evidence against the standard disk accretion scenario. 
The rapid spin-down torque of OAO 1657-415 could be explained by a retrograde disk scenario, while the non-correlation between torque and flux during slow spin-down/spin-up episodes could be explained by a significant contribution to the observed fluxes from quasi-spherical wind accretion. 
The mean pulsed fluxes of OAO 1657-415 during rapid spin-down episodes are smaller than those during spin-up episodes, and this fact is also similar to the behavior of GX 1+4 \citep{Chak1997}.

The orbital profiles of OAO 1657-415 during spin-up and spin-down episodes are quite different, with reduced fluxes around ingress and egress periods for spin-down episodes (Fig.~\ref{fig:eopf}). 
Such reduced fluxes may result from more material between the neutron star and the observer, as revealed by the smaller hardness ratios of spin-down episodes around ingress and egress periods (Fig.~\ref{fig:udhr}). 
These different orbital behaviors of OAO 1657-415 between spin-up and spin-down episodes clearly show that the spin behavior of the neutron star is connected with the material flow at the orbital scale, not only the disk-magnetosphere interactions near the neutron star, supporting the retrograde/prograde disk accretion scenario. 
The differences in pulse profiles between spin-up and rapid spin-down episodes indicate that the interaction between a retrograde disk and the magnetosphere may lead to 
some different flow and emission patterns near the neutron star. 

\citet{Bay1997} proposed that the torque reversals of OAO 1657-415 could be explained by alternate prograde and retrograde disks formed through flip-flop instabilities found in hydrodynamic simulations \citep{Fry1988,Blon1990}. 
However, the spin-up and spin-down periods of OAO 1657-415 are much longer than the binary orbital period ($\sim10.4$
days) and the flip-flop timescale, $\tau_{\rm fl}\sim R_{\rm acc}/v_{\rm rel}\sim10^4\,\rm s$,
where $R_{\rm acc}$ is the accretion radius and $v_{\rm rel}$ is the relative velocity of the neutron star with
respect to surrounding materials \citep{Blon1990}.

Recent wind accretion simulations performed by \citet{Xu2019} demonstrated that OAO 1657-415 could host a persistent disk-like structure due to large upstream gradient and orbital effect. 
As pointed out in previous studies \citep[e.g.][]{Chak1997,Nelson1997}, for systems of perfect Roche lobe overflow, the formation of a stable retrograde disk seems implausible. 
We note that the system of OAO 1657-415 is far from a perfect Roche lobe overflow. The eccentricity of OAO 1657-415 is
$\sim\,$0.1, not reaching a circular orbit yet.
On average, the radius of the optical star is about $24.8\pm1.5\,R_{\sun}$, which is smaller than the estimated Roche-Lobe radius of $27.6\pm1.6\,R_{\sun}$ \citep{Mason2012}. 
In addition, the spin period of the optical star may not be the same as the orbital period. 
OAO 1657-415 also has one of the slowest terminal wind velocities at $v_{\infty}\sim250\,\rm km\,s^{-1}$ \citep{Mason2012}.
Combining these conditions, formation of alternate retrograde/prograde disk-like structure stable for tens of days might be possible. 

To explain the spin-down torque of OAO 1657-415, \citet{Kim2017C} proposed that a magnetized wind is probably accreted through a non-Keplerian disk (magnetic levitating disk), the inner radius of which is 
larger than the standard magnetosphere. While the magnetic levitating disk scenario can provide a reasonable spin-down torque, it seems hard to explain the torque-dependent orbital profile and hardness ratio profile of OAO 1657-415.

A preliminary analysis of Vela X-1 and Cen X-3 shows that their orbital profiles are also different for different torque states, although without a clear correlation between $\dot{\nu}$ and flux. 
It is possible that for Vela X-1 and Cen X-3, the accreting flow contains alternating angular momentum on a long timescale, but the luminosity is dominated by a quasi-spherical accretion.
The observed distribution of $\dot{\nu}$ vs flux of OAO 1657-415 is scattered, which is unlike the prediction of a pure disk accretion model and implies contamination of wind accretion or that the accreting structure is just ring-like or torus-like.
It is interesting to note that recently \citet{Sharma2021} discussed possible spin variations of OAO 1657-415 on a timescale of $\sim$5\,years \citep[see also][]{Barn2008} and they noted the torque reversal around 2015, which may be related with variations of stellar wind. In principle, the variation of stellar wind could also be a possible reason for the torque-dependent orbital profiles reported here.
Detailed hydrodynamic simulations with configurations similar to OAO 1657-415 are needed to test whether alternate
retrograde/prograde disk-like/torus-like structure stable for tens of days are possible. 
If confirmed, the prograde/retrograde scenario may also be applicable to other sources and could have important implications for the spin evolution of X-ray pulsars.

\section*{Acknowledgements}
We thank our referee for detailed comments that improved the paper much
and Sean Lake, Long Ji, Kun Xu and Yufeng Li for helpful discussions.
JL acknowledges the support by National Natural Science Foundation of China (NSFC, 11773035, U1938113) and by the Scholar Program of Beijing Academy of Science and Technology (DZ BS202002). 
LG acknowledges the support by the National Program on Key Research and Development Project (2016YFA0400804), and by the NSFC (U1838114), and by the Strategic Priority Research Program of the Chinese Academy of Sciences (XDB23040100).

\section*{Data availability}

The data utilized in this article are observed by {\it Fermi}/GBM, {\it Swift}/BAT and {\it MAXI}/GSC, which are publicly available at
\newline
\href{https://gammaray.nsstc.nasa.gov/gbm/science/pulsars.html}{https://gammaray.nsstc.nasa.gov/gbm/science/pulsars.html}, 
\href{https://swift.gsfc.nasa.gov/results/transients/BAT\_current.html}{https://swift.gsfc.nasa.gov/results/transients/BAT\_current.html}, \href{http://maxi.riken.jp/star\_data/J1700-416/J1700-416.html}{http://maxi.riken.jp/star\_data/J1700-416/J1700-416.html}.



\bibliographystyle{mnras}
\bibliography{oao1657}





\bsp	
\label{lastpage}
\end{document}